\documentclass[english]{sobraep}

\usepackage{ragged2e}

\title{NeuroCADR: Drug Repurposing to Reveal Novel Anti-Epileptic Drug Candidates Through an Integrated Computational Approach}

\author{Srilekha Mamidala\\
	\normalsize Garnet Valley, Pennsylvania, United States of America\\
	\normalsize E-mail: srilekha@wharton.upenn.edu
}

\begin{document}

\maketitle

\begin{abstract}
	Drug repurposing is an emerging approach for drug discovery involving the reassignment of existing drugs for novel purposes. An alternative to the traditional de novo process of drug development, repurposed drugs are faster, cheaper, and less failure prone than drugs developed from traditional methods. Recently, drug repurposing has been performed in silico, in which databases of drugs and chemical information are used to determine interactions between target proteins and drug molecules to identify potential drug candidates. A proposed algorithm is NeuroCADR, a novel system for drug repurposing via a multi-pronged approach consisting of k-nearest neighbor algorithms (KNN), random forest classification, and decision trees. Data was sourced from several databases consisting of interactions between diseases, symptoms, genes, and affiliated drug molecules, which were then compiled into datasets expressed in binary. The proposed method displayed a high level of accuracy, outperforming nearly all in silico approaches. NeuroCADR was performed on epilepsy, a condition characterized by seizures, periods of time with bursts of uncontrolled electrical activity in brain cells. Existing drugs for epilepsy can be ineffective and expensive, revealing a need for new antiepileptic drugs. NeuroCADR identified novel drug candidates for epilepsy that can be further approved through clinical trials. The algorithm has the potential to determine possible drug combinations to prescribe a patient based on a patient’s prior medical history. This project examines NeuroCADR, a novel approach to computational drug repurposing capable of revealing potential drug candidates in neurological diseases such as epilepsy.
\end{abstract}




\section{INTRODUCTION}

The traditional method of developing drugs is time-consuming, expensive, and has substantial risk of failure. The process has five main stages: preclinical trials or discovery, preclinical trials, clinical research, FDA review, and post-market FDA safety monitoring [1]. 

\subsection{Traditional Drug Development Process}

In the discovery stage, drug candidates are established by researchers through multiple avenues. The most significant of these are novel insights concerning a disease, allowing researchers to develop ways to target these aspects. In addition, researchers can run tests of molecular compounds [2] against many diseases to determine if any drugs show promise for potential treatment. The discovery stage consists of thousands of drug candidates that are then reduced to very few for preclinical trials. 

The preclinical trials stage is conducted via two methods, in vitro or in vivo, to determine the toxicity of a drug candidate, in addition to providing information on appropriate dosing [2]. Combined, the discovery and preclinical trials stages are three to six years long and cost \$5 to 7 million [1].

	Clinical research constitutes the bulk of the traditional drug discovery timeline. This stage aims to measure the interaction and effect of a drug on the human body. To design a clinical study, researchers determine the selection criteria for participants and the research protocol. These trials often range from smaller, Phase 1 studies to larger scale, Phase 3 studies [2]. As most clinical studies occur over several months, they lengthen the amount of time needed to bring a drug to market, while also raising the cost of a drug due to the heavy expenses involved in designing robust trials. The clinical phase takes between six to eight years, and costs anywhere between \$20 to 40 million for a single drug across all three phases [1]. Furthermore, many drugs do not pass clinical trials for a variety of reasons, ranging from lack of funding to insufficient understanding of drug interactions [1]. 
 
	Drugs that pass these trials are then sent to be certified by the Food and Drug Administration (FDA). Information supplied for certification ranges from data from all trials to directions for use and is intended to provide a complete perspective of the drug, resulting in a drug’s labeling for market [2]. The FDA approval process can take 6-10 months, furthering a drug’s introduction to market [1].  
 
	Finally, the FDA conducts post-market safety monitoring during a drug’s use in the market. Monitoring of a drug often centers around reviewing safety reports to accordingly adjust dosage rates and information for proper usage [2]. 
 
	Developing a new drug with this method can cost millions, and even billions of dollars and take more than a decade to develop [1]. With many drugs failing at the early stages of development, patients with certain conditions are forced to have unaffordable drug treatments [3] due to the high prices that drugs are listed for on the market. For some diseases, patients may have virtually no drug medications, with few options left for other types of treatment.

\subsection{Drug Repurposing}

One emerging alternative to the traditional drug development process is drug repurposing. Drug repurposing involves the reassignment of existing drugs for novel therapeutic purposes [1] [4] [5] [6]. The logistics of drug repurposing can be understood by the fundamental action of a drug. Drugs work by interacting with receptors on cell surfaces or enzymes. These molecules abide by the “lock and key” model in which they are constructed as specific three-dimensional structures that molecules need to fit exactly into for a successful interaction. 

Drugs function as either agonists or antagonists, meaning that they either mimic or prevent a molecule from attaching to a receptor [2]. Therefore, a drug can either replace a deficient molecule’s activity or prevent a harmful interaction from occurring. Drug repurposing uses this concept because diseases often target multiple receptors in the body. Therefore, malfunction of a receptor can signal the onset of potentially multiple diseases. If a drug that targets a certain receptor area, which is also a target of another disease, this drug may be a candidate for the latter disease, depending on many other attributes of the drug such as structure and other molecular interactions. 

Drugs that are repurposed can be approved, discontinued, or in the process of clinical trials [2]. This process has several advantages over the traditional method of drug discovery. They are billions of dollars cheaper than traditionally created drugs, in addition to being much faster to develop and send to patients [5]. These types of drugs are also less failure prone than traditional drugs as many have already passed clinical trials and exist in the market, indicating their nontoxic use in humans. 

The COVID-19 pandemic has accelerated the need for repurposed drug treatments, as medications to treat patients needed to be developed quickly as well as robustly to account for multiple variants of the virus. The lengthy timeline required by the traditional process has proved insufficient, in addition to the high failure rate that results. For instance, repurposing studies performed in Marseille have shown that hydroxychloroquine may contain useful molecules to combat COVID-19 [7]. Drug repurposing has allowed treatments to be pushed to market faster and with more confidence in the intended interactions with patients, therefore saving millions of lives.

\subsection{Approaches to Drug Repurposing} 

There are multiple approaches to drug repurposing. One is an experimental approach where drugs are screened to evaluate their effectiveness in treating conditions via pharmacological assays [1]. A subset of the experimental approach is the clinical approach in which patients with a certain condition are given potential drug candidates, selected by analysis of patient tissue or blood [4]. The clinical approach necessitates fewer resources compared to traditional clinical trials because interactions of the drug with the human body are already known, lowering the amount of risk involved [1] [4]. In vitro and in vivo disease models are also used to determine drug molecules that may aid in disease treatment [1]. 

Repurposed drugs have also been discovered serendipitously throughout medical history [1]. The first instances of repurposed drugs were discovered through this method. However, serendipitous discovery is not a true method of drug discovery as it fails to be consistent. 

Drug repurposing has also recently been performed via an in silico approach, in which databases of drugs and chemical information are used to identify potential drug candidates by forming associations between drug structures and protein and genes [5] [8] [9] [10]. In silico approaches have been gaining popularity due to the increased accessibility to drug molecule databases. Potential drug molecules are identified by analysis of interactions between the drug and disease targets [5]. Computational approaches have been shown to be time and labor efficient [8] [9] [10] enabling many drug candidates to be returned that can then be verified by further experimental investigation as shown in Figure 1.

\begin{center}
  \includegraphics[width=0.45 \textwidth]{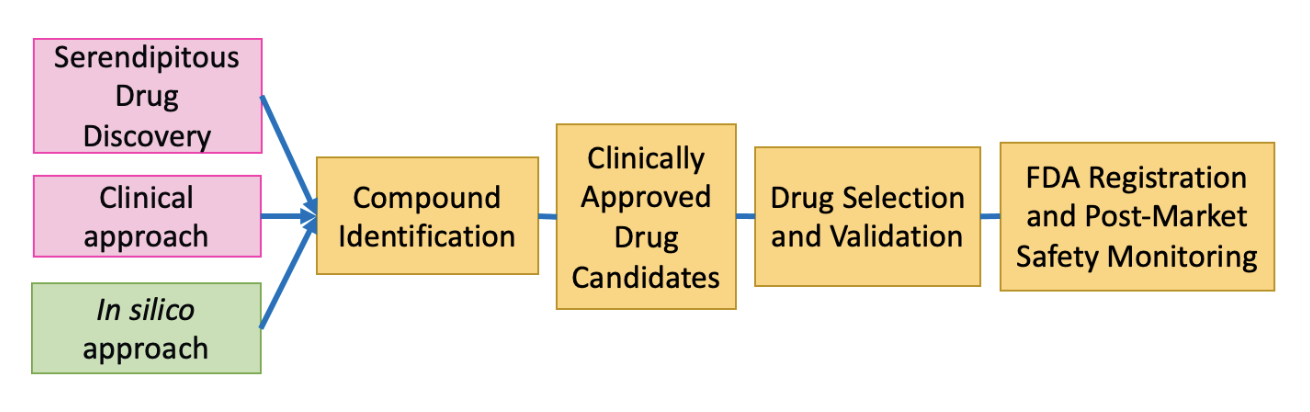}
\end{center}

\centering \emph{Figure 1: The drug repositioning process}

\subsection{Drug Repurposing in Epilepsy} 

\justifying

One condition that could greatly benefit from drug repurposing is epilepsy. Epilepsy is a condition characterized by seizures, periods of time with bursts of uncontrolled electrical activity in brain cells15. Normally, messages travel through the brain in an orderly sequence of electrical activity. When neurons fire to transmit a signal, they undergo several processes, the most relevant of which are depolarization and repolarization. In depolarization, the neuron will open sodium channels to allow sodium ions to enter the neuron, causing the charge inside to become positive and conduct electrical signals. When the neuron has stopped firing, these gate close and potassium ions that are inside the neuron are guided out through potassium channels, restoring the negative charge of the neuron in a process known as repolarization. In this cycle, the nerve can control firing periods and rest periods accordingly. In patients with epilepsy, however, the sodium gated channels become dysfunctional during depolarization, allowing too much sodium to enter the neuron resulting in heightened excitation. In addition, during repolarization, potassium channels do not allow as much potassium to exit the neuron [4] [11] causing neural firing to persist resulting in uncontrolled electrical activity in the brain. 

\begin{center}
  \includegraphics[width=0.45 \textwidth]{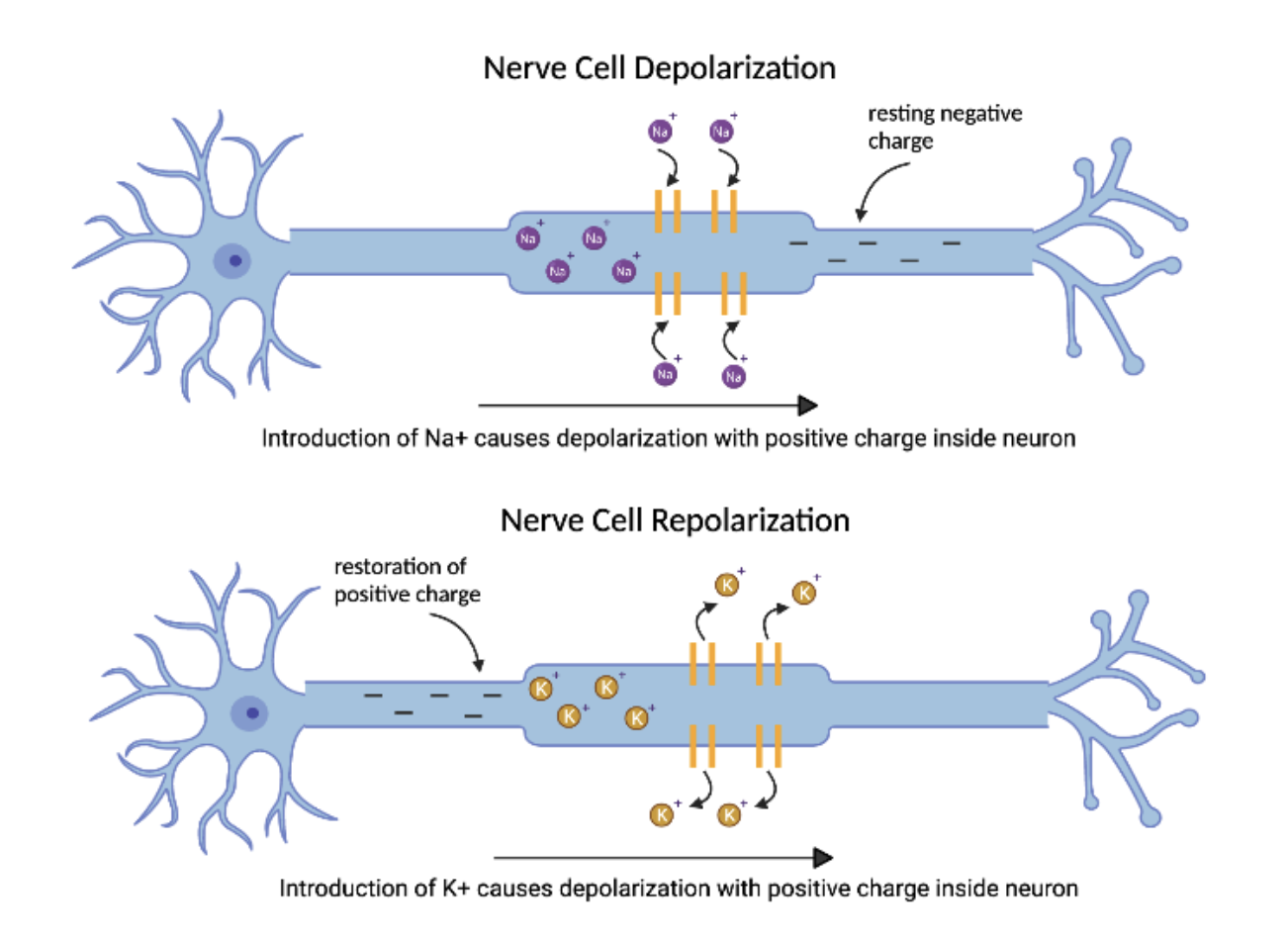}
\end{center}
\centering \emph{Figure 2: Processes of nerve depolarization and repolarization in patients with epilepsy}

\justifying

\vspace{5mm} 

Epilepsy is an umbrella term for many types of seizures, as some types of seizures target only specific parts of the brain. Called focal seizures, these are reflected in the symptoms that are experienced by the patient [11]. Due to this large variation, not all types of seizures can be treated with anti-epileptic drugs. Anti-epileptic drugs function mainly by regulating the aforementioned ion channels in neurons to regulate firing by either decreasing excitation or increasing inhibition [4]. Furthermore, existing anti-epileptic drugs can have many side effects, revealing a need for new treatment. Current anti-epileptic drugs on the market include sodium valproate, levetiracetam, and eslicarbazepine acetate, all of which report side effects such as lack of energy and agitation [12]. Finally, anti-epileptic drugs can also be expensive depending on the adverse effects of the drugs. Therefore, patients are forced to sacrifice either their health or money for a medication essential for survival. This harsh situation faced by patients with epilepsy can potentially be reduced through drug repurposing. 

It was hypothesized that combining several machine learning approaches, namely decision trees, random forest regression, and k-nearest neighbor (KNN) algorithms, would result in a greater number of repurposed drug candidates and more accurate drug predictions as the combination of several approaches allows for classification that can be cross-checked among several algorithms, resulting in greater functionality. 

The combination of these techniques eventually became NeuroCADR - a novel computational platform for drug repurposing. NeuroCADR can identify novel drug candidates for many different diseases that can be further approved through clinical trials and eventually used to treat patients. This paper will discuss the specific application of NeuroCADR to identify potential medications to treat epilepsy. Furthermore, the NeuroCADR platform can potentially be incorporated into a user-friendly website that medical professionals can use to determine possible drug combinations to prescribe a patient based on a patient’s prior medical history, the implementation of which will be discussed later in this paper. 


\section{METHODS}

\begin{center}
    \includegraphics[width=0.45 \textwidth]{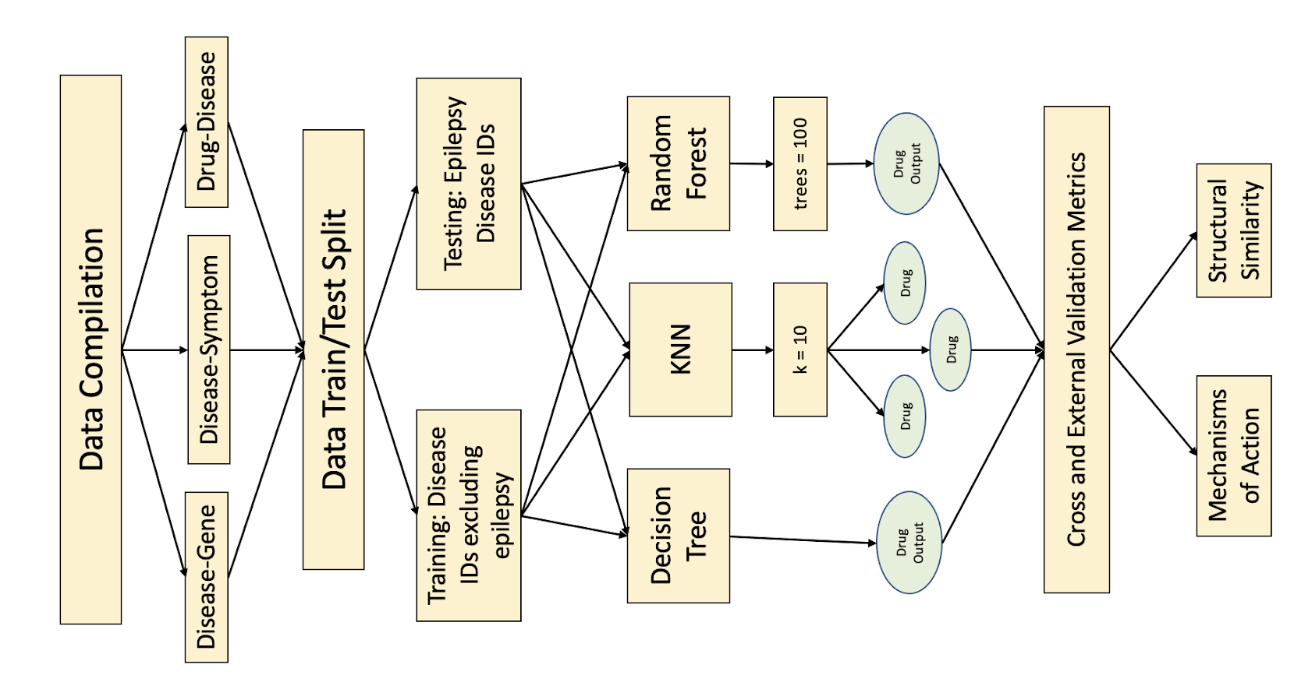}
\end{center}

\centering \emph{Figure 3: Proposed approach for returning of epileptic drug candidates}

\justifying

\subsection{Data Compilation} 

Drug repurposing has been able to be performed computationally due to the availability of large libraries such as DrugBank [12]. All datasets used in this study have been verified and used in numerous other scientific literature. Data was compiled by first determining which attributes of the drug were most important to include in the final dataset. 

It was determined that the final dataset to be used for training should be disease-centered. This was chosen in opposition to a drug-centered approach due to the fact that the NeuroCADR platform runs based on an initial input of disease to return potential drug candidates. Therefore, the chosen approach is optimal because it is easier to extract disease profiles so that drug candidates can be directly compared to associated symptoms of a disease. On the other hand, a drug-centered approach would cause confusion due to the fact that some drugs are already listed for multiple diseases (they are already repurposed). 

From this organizational standpoint, it was determined that the “master” dataset should contain the disease itself, drugs affiliated with the disease, symptoms of the disease, and genes associated with the disease and therefore associated with connected drugs. 

\subsubsection{Drug-Disease Dataset} 

In order to map drugs to diseases, data from the Comparative Toxicogenomics Database (CTD) containing chemical-disease associations was used, consisting of 466,657 disease entries [13]. The CTD database expressed diseases using the MeSH ID, chosen for its comprehensive identification of conditions across many disease categories. Drugs were expressed using DrugBank IDs as DrugBank provides the most complete list of drugs and substances. Abbreviated IDs were chosen in place of the full names of the entries to allow for easier sorting and data extraction. 

\subsubsection{Disease-Symptom Dataset}

To map the disease IDs to their respective symptoms, a dataset was taken from the Human Symptoms-Disease Network [14]. Data was organized alphabetically by symptoms, ranging from premature aging to eye pain. Symptoms were also represented with a symptom ID to ensure compactness of the data. This dataset contained 21,177 entries. Each disease that was affiliated with a certain symptom was represented in its own row, such that one disease could appear in multiple rows. Data represented diseases in the MeSH ID, which was necessary as the next step consisted of mapping diseases in the drug-disease dataset to those of the symptom dataset, essentially combining the two. This was done by first determining the unique number of symptoms in the dataset. These unique symptom IDs were extracted and entered into a list, which were then appended as columns into the Drug-Disease dataset that was created beforehand. This enabled the dataset to be expressed in binary - data points that existed, such as a disease having a certain symptom, were represented by a one, while nonexistent data points were represented as zeroes. However, in order to combine the two datasets, each disease in the Drug-Disease dataset needed to have at least one symptom in the Disease-Symptom Dataset. Since this was not true for all diseases, each disease in the Drug-Disease dataset was cross checked to see if an entry for it existed in the latter dataset, and if such an entry did not exist, the disease was dropped.

\subsubsection{Disease-Gene Dataset}

The Disease-Gene associations were taken from DisGeNET [15]. Genes that are connected to a disease are affiliated with respective drug treatments as drugs target certain receptors whose structures are encoded by genes. Data consisted of the gene symbol, gene ID, the disease name, and disease ID and contained 1,048,547 genes. The dataset was grouped by gene, rather than disease, meaning that extra sorting needed to be done to add the data into the dataset. The method of extracting useful information is similar to that of 2.1.2. Unique genes were collected and appended as columns to the complete dataset, with one representing an interaction and zero representing no interaction. Similar to the previous data integration, IDs were used rather than gene names to allow for easier reading.

\begin{center}
    \includegraphics[width=0.5 \textwidth]{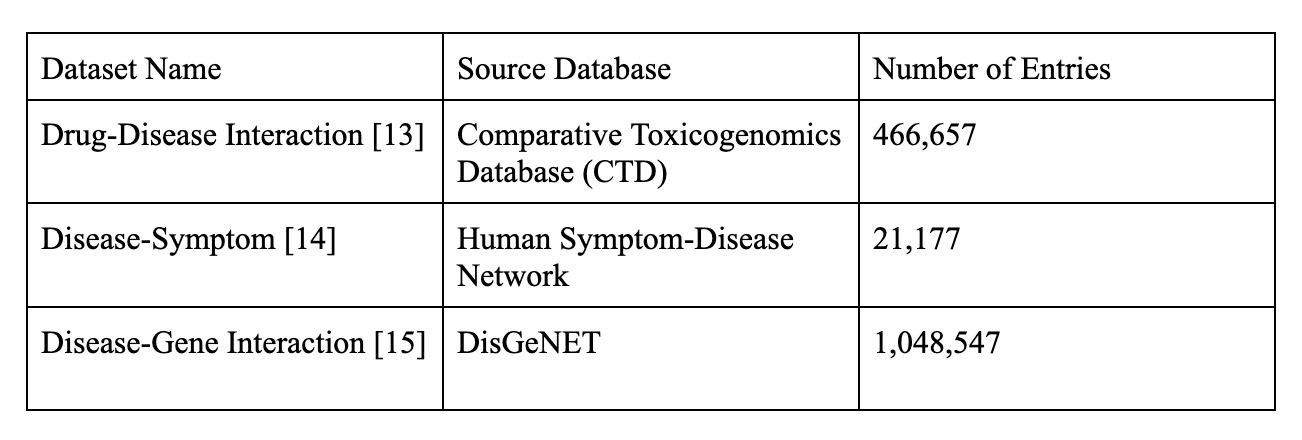}
\end{center}

\justifying

\subsubsection{Drug-Disease-Symptom-Gene Dataset}

The resulting dataset consisted of disease IDs mapped to their respective existing drug treatments and gene affiliations, with 81,744 diseases cataloged. This “master” dataset was then used for training of the algorithm. 

\subsection{Algorithm Construction}

The NeuroCADR platform consisted of a multi-faceted approach of multiple machine learning algorithms in order to provide a comprehensive list of drug candidates, computed via distinct methodologies to provide a robust list. The three machine learning approaches in use were k-nearest neighbors (KNN), decision trees, and random forest regression. 

KNN was chosen as it requires no training time [5], contrary to deep learning and other neural network types (CNN, RNN). KNN algorithms are relatively straightforward and only require tuning one parameter at a time, the value of k. This singular tuning makes establishing associations between drugs and drug targets more streamlined and simpler to visualize. KNN is a supervised machine learning algorithm, requiring input data to analyze patterns and predict output data when given new unlabeled data. KNNs are also shown to have greater theoretical guarantees than other similar algorithms [16]. 
The primary assumption in a KNN algorithm is that similar data exists near each other. Therefore, associations can be made by finding the distance between two points with a given number of “neighbors”, denoted as k. The selected k-value produces the least number of errors with the training data while still being able to make accurate predictions. In addition, KNNs use the concept of a k-fold cross validation. K-fold cross validation is a way to evaluate algorithms with new data by dividing it into k shuffled groups. In each of k iterations, one portion is set aside while the others are trained. The separated portion is then used as testing data [8]. 

Decision trees operate through a tree-like model in which data is fed through filters that make decisions, resulting in a series of consequences that affect the final outcome of an event [17]. This approach was chosen to contrast the KNN method, which returns a list of possible drug candidates. However, the decision tree method returns only the best drug candidate selected via subsequent levels of smaller decisions. The returning of only one drug can help provide insight into the accuracy of the algorithm itself and the completeness of the dataset as well. Decision trees are a type of supervised machine learning and require little data preparation as well, reducing run time [17]. 

Random forest regression is also a supervised machine learning algorithm using an ensemble method (combines multiple machine learning algorithms) to limit dependency on a single model and therefore return a more accurate result. The random forest regression works somewhat like that of the decision tree, except each tree runs as its own prediction model and these decisions are then averaged or “voted” upon to return the final result. The changeable parameter in the random forest model is the number of trees, or estimators, that are used [18]. A greater number of trees may signal a more accurate output, but needs to be checked for potential overfitting. The random forest model was chosen as it takes different models into account, which is the essence of the NeuroCADR platform.
\newline

\subsection{Algorithm Testing and Training}

Parameters for training were then defined. Features for training were the “drug”, “disease”, “symptom”, and “gene”. The k-value for the KNN algorithm was chosen to be 10, representing the nearest number of drugs to check for to assign a repurposing score. The k-value was selected by running the algorithm with different values of k to find the optimal value. The value producing the least number of errors with the training data along while still being able to make accurate predictions was chosen. The number of estimators was chosen similarly and was set to 100. The number of folds for cross validation was set to 10, and the analysis was set to run 10 times. A GitHub repository containing all data and code is linked here.  

The dataset was read in and any potential duplicate entries were dropped to prevent possible overfitting. The disease ID for generalized epilepsy was selected [12], as well as a secondary form. On the NeuroCADR platform, these IDs would be inputted by the user in order to generate drug candidates for another disease. The complete dataset was then split into training and testing data. Rather than the usual random split of data, testing data was designated to only contain entries the disease ID for generalized epilepsy and its related counterpart. This was done due to the fact that if entries for epilepsy were included, the algorithm would simply return these entries rather than selecting other drug candidates. 

The model was then prepared and run through the respective models (KNN, decision tree, random forest regression). Testing and returning of the drug candidates occurred through the input of the data entries that only contained the epilepsy disease IDs. 

\section{RESULTS AND DISCUSSION}

\subsection{Drug Candidate Outputs}

NeuroCADR found multiple potential drug candidates based on the datasets inputted. Accuracy of the drug candidates can be measured by structural similarity to drugs that are currently being used to treat epilepsy in addition to cross and external validation metrics. 

The NeuroCADR platform outputted several drug candidates to be used for epilepsy. 

\subsubsection{Decision Tree Model}

When run in the decision tree, the drug cetrorelix (DB00050) was outputted. Cetrorelix is an antagonist of a GnRH, gonadotropin releasing hormone and is used to prevent hormone surges and the premature releasing of eggs in women during hormonal and reproductive treatment [12]. Cetrorelix does this by having GnRH control the luteinizing hormone, which starts ovulation during the menstrual cycle. By blocking the release of this hormone, the release of eggs can be inhibited [12]. It was hypothesized that cetrorelix was selected as a candidate because malfunction of GnRH is known to potentially be caused by epileptic seizures [19]. Therefore, a drug that targets this hormone in order to stabilize it may act upon anti-epileptic mechanisms as well. A more specific reasoning as to the reason for selection of cetrorelix is unknown and is certainly a future topic of interest.

\subsubsection{Random Forest Model}

When run in the random forest regression, an unexpected result occurred. The drug that was returned was valproic acid (DB00313). This result was unexpected as valproic acid is an existing seizure treatment, signaling a potential inconsistency in the dataset. However, if valproic acid was listed as a treatment for a disease other than epilepsy, then it may signal the accuracy of the algorithm as the correct attributes were matched to return the drug. Valproic acid was originally used as an organic solvent until George Carraz serendipitously discovered in 1963 [12] that it could be used to prevent epileptic activity as an anticonvulsant. In addition, valproic acid is being used for treatment against migraines and possibly in oncology [12]. 

\subsubsection{KNN Model}

Multiple drugs were returned by the KNN algorithm. Drugs were returned in order of relevance to epilepsy treatment. The most relevant drugs returned were 1) thiamylal (DB01154) 2) metronidazole (DB00916) 3) alitretinoin (DB00523) 4) dazoxiben (DB03052) and 5) malachite green (DB03895). Other drugs that were returned included nicotinamide (DB02701) and bleomycin (DB00290). 

Thiamylal is a drug molecule classified as a barbituate that is prescribed for inducing a short anesthesia or hypnotic state [12]. In addition, it is also sometimes combined with common painkillers such as acetaminophen to introduce sedative effects [12]. It was hypothesized that thiamylal was selected as a drug candidate due to its mechanism of action, which involves the neurotransmitter GABA (gamma aminobutyric acid). Reduced levels of GABA have been connected to seizure activity [20]. Thiamylal operates by binding to a Cl- ionophore at a GABA receptor and increasing the amount of time this ionophore is open, therefore increasing the inhibitory effect of GABA [12]. 

Metronidazole is part of a group of antibiotics called nitroimidazoles, used to treat a variety of infections ranging from bacterial infections to inflammatory lesions of rosacea [12]. The antiparasitic properties of this drug molecule has made it applicable to treat this wide range. Metronidazole has also been used, off-label, for treatment of Crohn’s disease [12]. It is unknown as to why metronidazole was chosen as a drug candidate for epilepsy. It is known that this drug exhibits inhibitory activity in certain types of DNA [12], and this may counteract the uncontrolled neuron signals in the brains of patients of epilepsy. 

Alitretinoin is used to treat Kaposi’s sarcoma along with eczema and other skin conditions off-label as a vitamin A derivative [12]. Alitretonin operates in the body by binding to all intracellular retinoid receptor subtypes such as RARa, RARb, RXRa, RXRb, and RXRg [12]. These receptors then regulate the expression of genes and control processes such as cellular differentiation [1]. There is no existing scientific literature mentioning alitretinoin as a potential treatment for epilepsy. However, retinoic acid, a related molecule that is also a vitamin A metabolite23 and interacts with the same receptors, has been shown to exhibit “antiepileptogenic effects” through the “modulation of gap junctions, neurotransmitters, long-term potentiation, calcium channels and some genes” [1]. Therefore the similar activity of alitretinoin may introduce these same anti-epileptic effects. 

Dazoxiben is an organic compound belonging to the benzoic acid class and is a “orally active thromboxane synthetase inhibitor” [12] used in the treatment of Raynaud’s syndrome, in addition to pulmonary hypertension treatment [21]. The exact mechanism of action for dazoxiben is unknown, but this molecule is an enzyme inhibitor [21] [22]. Thromboxane inhibition has been linked to prevention of seizures through a study that researched the effect of COX-2 inhibition on epileptic seizures [23]. COX-2 is an enzyme that is rapidly produced in large amounts during seizures and increased levels in certain areas of the brain during seizures has been reported. COX-2 catalyzes the process of converting arachidonic acid to PGH-2, which then converts to five prostanoids, one of which is thromboxane [23]. Therefore, this relation between COX-2 and thromboxane may be the reason why dazoxiben was chosen as a drug candidate. 

Malachite green is an organic chloride salt that is used as a dye or as an antifungal agent in aquaculture [24]. It is unknown as to why malachite green was chosen as a drug candidate. However, it is probable that malachite green is not a practical candidate due to its numerous side effects and its classification as a carcinogen [24]. 

The performance of NeuroCADR was compared to several other existing computational drug repurposing methods such as logistic regression, which was unable to even parse through the large data to make meaningful conclusions. The proposed algorithm matched closest in performance to deep learning which was expected as KNN is a subset of deep learning and neural networks [16] and therefore follows the same general principles. The accuracy of NeuroCADR can be seen through the relevance to epilepsy that the returned drug candidates exhibited. 

The success of NeuroCADR was compared to that of a clinical approach to epilepsy, where hippocampal brain tissue of patients with epilepsy was analyzed with RNA sequencing [4]. NeuroCADR was able to identify a greater number of potential drug candidates. The above study also tested the effectiveness of the most promising drug candidates on zebrafish [4] and concluded positive results, showing one advantage of a clinical approach.

\subsection{Potential Limitations}

One possible limitation of this algorithm is the data that was used as the platform only analyzed drugs that are approved. This error can be mitigated by further training of the algorithm using drugs that are in later stages of clinical trials, for example. 

In addition, overfitting or underfitting of the data may have occurred. Overfitting the data would have caused “false positives”, drug candidates that are realistically not suitable for treatment for epilepsy, while underfitting the data would have caused certain drugs that may be practical for treatment to not be recorded by the algorithm. 

Parameters such as the value of k in the KNN algorithm and the number of estimators in the random forest model were determined experimentally. Changing this value may yield a slightly different list of drug candidates. 

\subsection{NeuroCADR Platform}

The NeuroCADR platform could potentially be implemented into a website that can be used by doctors and other medical professionals to reveal potential drug candidates to prescribe patients based on their prior medical history. 
This website is projected to include sections describing the algorithms used and the concept of drug repurposing itself. The platform operates by first allowing the user to select a disease that they want to visualize potential drug candidates for. Next, the user inputs other drugs that a certain patient may be taking. This is recorded so that any drugs that are returned by the platform will be flagged if they contain an interaction with the inputted drug. The NeuroCADR platform is not meant to be a medication tool for doctors to prescribe patients, but rather a tool to determine viable drug candidates that can be further verified through clinical studies and certified by the FDA. 

\begin{center}
   \includegraphics[width=0.5 \textwidth]{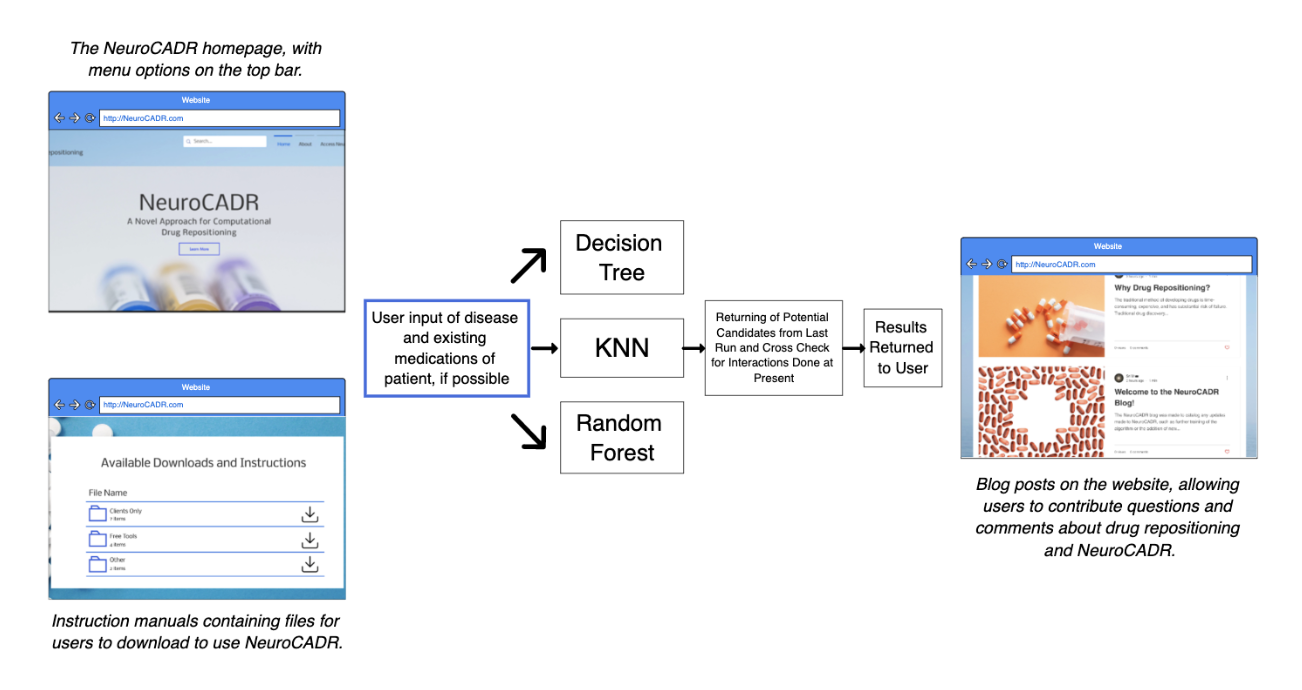}
\end{center}

\centering \emph{Figure 4: Wireframe of NeuroCADR website}

\justifying

\vspace{5mm} 

One principal design feature of the NeuroCADR website is the functionality of the platform. During the running of the algorithms, it was noticed that outputs would take an extended amount of time to be returned. This was declared inadequate as many users of the website would likely have computing devices incapable of reading the amount of data that is inputting into the platform. Therefore, it was proposed that in order to reduce user wait times and to handle multiple user requests more efficiently, the algorithm would run itself on regular intervals and return the result from the latest running. Experimentally, it was determined that the algorithm would run once every two hours to provide the most accurate results while preventing devices from becoming overwhelmed. This automated running process, as opposed to manual running, is advantageous in that it will automatically update results in the presence of new data and that it is more transferable across a multitude of devices as the platform itself is being run on a separate server.

\subsection{Impact and Applications}

\subsubsection{Development of Novel Pharmaceutical Treatments}

The NeuroCADR platform can be used by pharmaceutical companies to develop novel therapeutic treatments for patients with conditions that have minimal drug treatments. These companies can save billions of dollars per drug, in addition to being able to send drugs to the market in half the time of a traditionally developed drug. People with conditions that are being treated with repurposed drugs can have the opportunity to get affordable treatment in which the effects are already clearly known. 

\subsubsection{Establishment of Drug Repurposing Candidates for Other Diseases}   

NeuroCADR can be run to reveal novel drug candidates for diseases other than epilepsy, such as Parkinson’s disease, a severe neurodegenerative disorder that currently has no cure. In addition, orphan diseases, diseases that affect less than 200,000 people nationwide [3], would be greatly benefited by drug repurposing. Many orphan diseases currently do not have drugs developed for them due to the high financial cost needed to develop drugs via the traditional method, providing little financial incentive for pharmaceutical companies to develop them [3]. Drug repurposing can provide novel treatments for patients with these diseases due to the reduced cost involved.

\includegraphics[width=0.45 \textwidth]{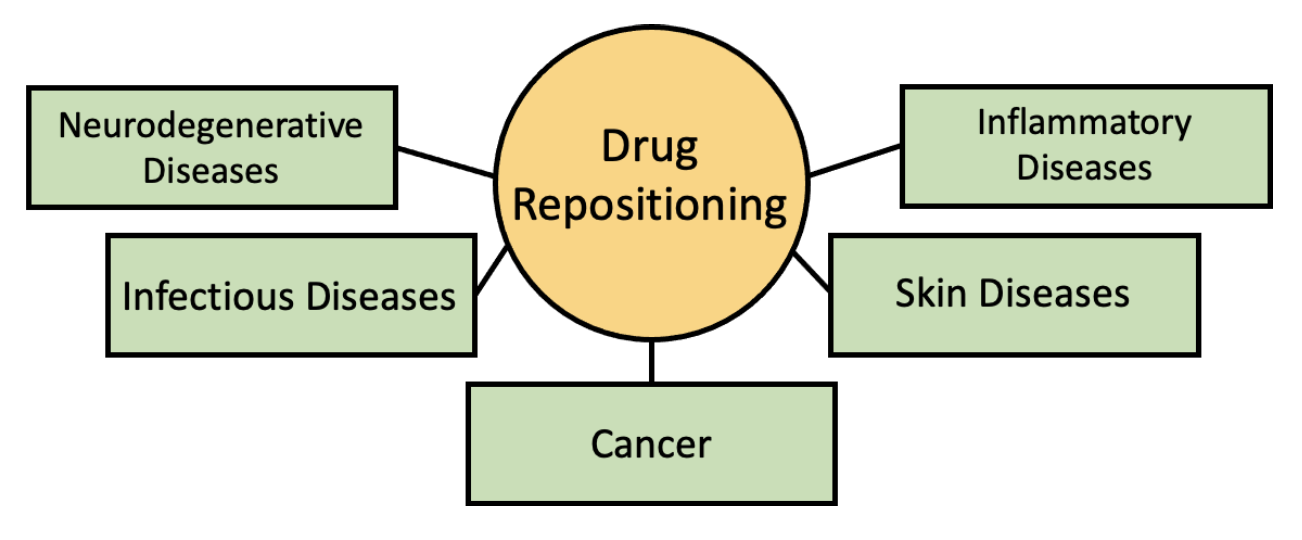}
\centering \emph{Figure 5: Disease groups treatable through drug repurposing}

\justifying

\subsubsection{Combating the Opioid Epidemic}

Opioids are a class of drugs that include legally prescribed drugs such as oxycodone as well as illegal drugs such as heroin and fentanyl. Opioids are mostly prescribed for pain relief and provide morphine-like effects, causing users to quickly become addicted to them [25]. Every year, 10.1 million people misuse prescription opioids. Of those people, over 70,000 died from a drug overdose. In addition, opioid overdose deaths have quadrupled since 1999 and are only increasing each year [25]. NeuroCADR can potentially be used to identify alternatives to opioid painkillers that can be prescribed to prevent drug addiction and drug overdoses. 

\subsubsection{Future Work}

The NeuroCADR platform can be expanded to to include different datasets of drugs, genes, and protein interactions to enable identification of potential drugs for a greater variety of diseases. Including drug profiles for discontinued drugs may also provide new insights on potential uses for these medications. 
Further model training and data classification for NeuroCADR will improve accuracy of drug candidates that are returned by the algorithm.

The use of a greater number of machine learning approaches may allow for greater cross-checking of drugs to boost confidence in the results outputted. The creation of a drug-centered dataset may also be considered to determine whether this would vary the drug candidates returned as the sorting of data proceeds differently. 

Developing the NeuroCADR website will allow for drug candidates to be further validated through clinical testing via pharmaceutical companies, paving the way for future repurposed drug treatments.

\section{CONCLUSIONS}

This project aimed to develop a novel computational approach for drug repurposing using a multi-faceted approach consisting of decision trees, random forest regression, and k-nearest neighbors to reveal potential drug candidates for epilepsy. The hypothesis that this algorithm would be more accurate than existing in silico methods was supported. NeuroCADR reported a greater number of drug candidates for epilepsy than other methods such as logistic regression and support vector machines. NeuroCADR also performed better than clinical approaches to drug repurposing by reporting a greater number of drug candidates.

The platform analyzed drugs using individual datasets containing associations between drug structures, associated diseases, and genes. Diseases were mapped to their respective drugs and symptoms to return a complete profile of the disease, performed by combing through each entry and dropping potential duplicates or inconsistencies. Next, the dataset was run through the multiple approaches to eventually return a list of potential drug candidates, ranked by computed relevance to the attributes of the disease. 

There are many applications of this project. NeuroCADR can help in the development of new pharmaceutical treatments for epilepsy by providing companies with information on the most plausible drugs to repurpose. The platform can also provide insight for treatments of other diseases that are often overlooked by pharmaceutical companies due to the high cost involved. The opioid epidemic is also an issue that NeuroCADR can assist with by providing less addicting alternatives to opioids. Drug repurposing as a whole has immense potential in the betterment of treatment development for many disease categories. 

\vspace{3mm} 

\textbf{Conflict of Interest}: The author has declared that there is no conflict of interest. 

\vspace{3mm} 

\textbf{Data Availability Statement}: All source data used is publicly available. Data used for training/testing can be provided upon request. 

\section{REFERENCES}

This project was performed with the guidance of Professor Jaudelice de Oliveira of Drexel University and PhD student Dubem Ezeh of Drexel University.

\RaggedRight

\vspace{3mm} 

[1] Rosiles-Abonce, A., Rubio, C., Taddei, E., Rosiles, D., \& Rubio-Osornio, M. (2021). Antiepileptogenic Effect of Retinoic Acid. Current neuropharmacology, 19(3), 383–391. https://doi.org/10.2174/1570159X18666200429232104 
[2] Rudrapal, M., Khairnar, S. J. , \& Jadhav, A. G.  (2020). Drug Repurposing (DR): An Emerging Approach in Drug Discovery. In  (Ed.), Drug Repurposing - Hypothesis, Molecular Aspects and Therapeutic Applications. IntechOpen. https://doi.org/10.5772/intechopen.93193 

[3] Marilyn J. Field and Thomas F. Boat, eds, Rare Diseases and Orphan Products: Accelerating Research and Development (Washington, DC: National Academy Press, 2010), 2.

[4] Brueggeman, L., Sturgeon, M. L., Martin, R. M., Grossbach, A. J., Nagahama, Y., Zhang, A., Howard, M. A., 3rd, Kawasaki, H., Wu, S., 

[5] Cornell, R. A., Michaelson, J. J., Bassuk, A. G. (2018). Drug repositioning in epilepsy reveals novel antiseizure candidates. Annals of clinical and translational neurology, 6(2), 295–309. https://doi.org/10.1002/acn3.703

[6] Jarada, T. N., Rokne, J. G., Alhajj, R. (2020). A review of computational drug repositioning: strategies, approaches, opportunities, challenges, and directions. Journal of cheminformatics, 12(1), 46. https://doi.org/10.1186/s13321-020-00450-7

[7] Xiangxiang Zeng, Siyi Zhu, Xiangrong Liu, Yadi Zhou, Ruth Nussinov, Feixiong Cheng, deepDR: a network-based deep learning approach to in silico drug repositioning, Bioinformatics, Volume 35, Issue 24, 15 December 2019, Pages 5191–5198, https://doi.org/10.1093/bioinformatics/btz418

[8] Colson, P., Rolain, J. M., Lagier, J. C., Brouqui, P., Raoult, D. (2020). Chloroquine and hydroxychloroquine as available weapons to fight COVID-19. International journal of antimicrobial agents, 55(4), 105932. https://doi.org/10.1016/j.ijantimicag.2020.105932

[9] Fahimian, G., Zahiri, J., Arab, S.S. et al. RepCOOL: computational drug repositioning via integrating heterogeneous biological networks. J Transl Med 18, 375 (2020). https://doi.org/10.1186/s12967-020-02541-3

[10] Park K. (2019). A review of computational drug repurposing. Translational and clinical pharmacology, 27(2), 59–63. https://doi.org/10.12793/tcp.2019.27.2.59

[11] Zhao, M., Yang, C. C. (2018). Drug Repositioning to Accelerate Drug Development Using Social Media Data: Computational Study on Parkinson Disease. Journal of medical Internet research, 20(10), e271. https://doi.org/10.2196/jmir.9646

[12] “Types of Seizures.” Centers for Disease Control and Prevention, Centers for Disease Control and Prevention, 30 Sept. 2020, https://www.cdc.gov/epilepsy/about/types-of-seizures.htm. 

[13] DrugBank Online | Database for Drug and Drug Target Info. (2022). Retrieved 27 June 2022, from https://go.drugbank.com/ 
The Comparative Toxicogenomics Database | CTD. (2022). Retrieved 28 June 2022, from http://ctdbase.org/

[14] Zhou, X., Menche, J., Barabási, AL. et al. Human symptoms–disease network. Nat Commun 5, 4212 (2014). https://doi.org/10.1038/ncomms5212

[15] DisGeNET - a database of gene-disease associations. (2022). Retrieved 28 June 2022, from https://www.disgenet.org/home/

[16] Guo, Gongde, Wang, Hui, Bell, David Bi, Yaxin. (2004). KNN Model-Based Approach in Classification. 

[17] Quinlan, J.R. Induction of decision trees. Mach Learn 1, 81–106 (1986). https://doi.org/10.1007/BF00116251
Breiman, L. Random Forests. Machine Learning 45, 5–32 (2001). https://doi.org/10.1023/A:1010933404324

[18] Fawley, Jessica A., Pouliot, Wendy A., Dudek, F. Edward Epilepsy and reproductive disorders: The role of the gonadotropin-releasing hormone network, Epilepsy and Behavior, Volume 8, Issue 3, 2006, Pages 477-482, ISSN 1525-5050, https://doi.org/10.1016/j.yebeh.2006.01.019.

[19] Khamsi, R. Warning on epilepsy drugs for young. Nature (2005). https://doi.org/10.1038/news051205-5 

[20] Belch, J. J., Cormie, J., Newman, P., McLaren, M., Barbenel, J., Capell, H., Leiberman, P., Forbes, C. D., Prentice, C. R. (1983). Dazoxiben, a thromboxane synthetase inhibitor, in the treatment of Raynaud's syndrome: a double-blind trial. British journal of clinical pharmacology, 15 Suppl 1(Suppl 1), 113S–116S. https://doi.org/10.1111/j.1365-2125.1983.tb02119.x 

[21] National Center for Biotechnology Information (2022). PubChem Compound Summary for CID 53001, Dazoxiben. Retrieved June 26, 2022 from https://pubchem.ncbi.nlm.nih.gov/compound/Dazoxiben. 

[22] Rojas, A., Jiang, J., Ganesh, T., Yang, M. S., Lelutiu, N., Gueorguieva, P., \& Dingledine, R. (2014). Cyclooxygenase-2 in epilepsy. Epilepsia, 55(1), 17–25. https://doi.org/10.1111/epi.12461 

[23] Malachite green. (2022). Retrieved 27 June 2022, from https://pubchem.ncbi.nlm.nih.gov/compound/Malachite-green

[24] “Understanding the Opioid Overdose Epidemic.” Centers for Disease Control and Prevention, Centers for Disease Control and Prevention, 1 June 2022, https://www.cdc.gov/opioids/basics/epidemic.html.

\end{document}